\documentstyle[11pt]{article}
\begin{document}
\centerline{\bf RESOURCE LETTER ON GEOMETRICAL RESULTS}

\centerline{\bf FOR EMBEDDINGS AND BRANES}

\bigskip\bigskip\bigskip

\centerline{{\bf Matej Pav\v si\v c}\footnote{\tt MATEJ.PAVSIC@IJS.SI}}

\bigskip

\centerline{\it Department of Theoretical Physics}

\centerline{\it Jo\v zef Stefan Institute}

\centerline{\it Jamova 39}

\centerline{\it SI--1000 Ljubljana, Slovenia}

\bigskip\bigskip

\centerline{{\bf Victor Tapia}\footnote{\tt
TAPIENS@CIENCIAS.CIENCIAS.UNAL.EDU.CO}}

\bigskip

\centerline{\it Departamento de F{\'\i}sica}

\centerline{\it Universidad Nacional}

\centerline{\it Bogot\'a, Colombia}

\bigskip\bigskip\bigskip

\centerline{\bf Abstract}

\bigskip

Due to the recent renewal in the interest for embedded surfaces we
provide a list of commented references of interest.

\bigskip\bigskip\bigskip

\centerline{\bf Introduction}

\bigskip

The idea that our space--time can be considered as a four--dimensional
space embedded in a higher--dimensional flat space is an old and
recursive one.  Recently, due to a proposal by Randall and Sundrum this
idea has again attracted much attention. Since several related ideas had
already been studied and several results had been obtained in the past we
have considered convenient to provide a full account of the existing
results on the subject.

The choice of the references included in this letter is selective and has
been dictated by some simple criteria. We have included only articles
published on wide circulation journals and books, etc., except when they
are of exceptional relevance or they has a historical character. 
We do not include electronic publications which have not passed a
refereeing procedure. A simple search at SPIRES gives more than 1200
references containing the words brane, membrane, etc. Other criterion for
inclusion in this letter is that the article made use of extrinsic
Riemannian geometry.

\bigskip\bigskip

\centerline{\bf A. The Geometry of Embeddings}

\bigskip

The concept of an abstract Riemannian manifold arises in mathematics as
the result of an evolution in mathematical attitudes. In the earlier
period, geometers thought of more concretely of curved surfaces embedded
in a flat Euclidean space (Gauss, 1827), {\it i.e.}, they  described the
geometry of surfaces as embedded in a higher--dimensional space. The
concept of an abstract Riemannian manifold, defined intrinsically, was
first explicitly formulated by Riemann in his famous, althought little
read, thesis, in 1850 (published only in 1868). In that work, what it is
today known as Riemannian geometry was introduced. The geometrical
properties of a manifold were characterised only by intrinsic properties,
without any need of a reference to a higher dimensional embedding space.

Almost inmediately after this abstract view came into favour, a question
naturally arose: the isometric embedding problem: The question of the
existence of concrete realisations of abstract Riemannian manifolds as
submanifolds of higher dimensional Euclidean spaces. Today we know that
the answer is yes: any intrinsically defined Riemannian manifold can be 
isometrically embedded, locally and globally, in an Euclidean space of
appropriate dimension and signature. The works cited in this section are
mainly oriented to establish this equivalence.

The embedding problem was first considered by Schl\"afli (1873) just
after the publication of Riemann work. He discussed the local form of the
embedding problem and he conjectured that a Riemannian manifold with
positive defined and analytic metric can be locally and isometrically
embedded as a submanifold of a Euclidean space $E_N$ with $N=n(n+1)/2$.

In 1926 Janet described a method of proof based on a power series
development, so it was limited to local results. Furthermore, he required
the metric to be analytic. This proof however, as Janet himself noticed,
was incomplete. He solved only the local problem for two--dimensional
manifolds with analytic metric. In 1927 E. Cartan extended the Janet's
proof to $n$--dimensional manifolds treating it as an application of his
theory of Pfaffian forms. The dimensionality requirement was
$N=n(n+1)/2$, as conjectured by Schl\"afli.

In 1931 Burstin completed the Janet's proof and also extended it to the
case in which the embedding space is a given Riemannian manifold $V_N$
with positive defined and analytic metric. The Gauss--Codazzi--Ricci
equations are conditions to be satisfied by the embedding, therefore,
they can be considered as integrability conditions for the embedding. In
1956 Leichtweiss gave a new proof based more substantially than Burstin
in the Gauss--Codazzi--Ricci equations of Riemannian geometry. His proof
is more involved than that provided  by Burstin. In 1961 Friedman
extended the theorem to Riemannian manifolds with indefinite metrics,
such as the space--time used in contemporary gravitational theories. For
semi--positive definite metrics see (Lense, 1926).

The first global isometric embedding theorem of $V_n$ into $E_N$ were
established by Nash (1956). The results depend crucially on the
compactness of $V_n$. For $V_n$ compact he obtains $N=n(3n+11)/2$; for
non--compact manifolds $N=n(n+1)(3n+11)/2$. The first global results for
indefinite metrics were obtained by Clarke (1970) and Greene (1970).
Recently, an improvement on Nash theorem was presented by Gunther (1989).

We may therefore conclude that any intrinsically defined Riemannian
manifold has a local and a global, isometric embedding in some Euclidean
space. Then one can consider the two approches, intrinsic and extrinsic,
to Riemannian geometry as completely equivalent.

This result opened new perspectives in theoretical physics, mainly in the
physics of the gravitational field. On the one hand it was considered as
a mathematical tool to construct and classify solutions of General
Relativity. On the other hand, it allows to introduce new variables,
perhaps more amenable for a description of quantum mechanical phenomena.

It is clear that some manifolds can be embedded in a higher dimensional
space with a dimension lesser tahn the requirement fixed by the embedding
theorem. Therefore, it is convenient to introduce the concept of class of
the embedding, which is the minimal number of extra dimensions necessary
to satisfy the Gauss--Codazzi--Ricci equations.

\bigskip\bigskip

\centerline{\bf B. Applications to General Relativity}

\bigskip

At the beggining of the {\tt XX} century General Relativity was developed
(by Einstein) based on Riemannian geometry. Almost inmediately exact
solutions were obtained based on the use of extrinsic geometry (Kasner,
1921). Several other results are contained in the references. A
particularly important result is that
Fridman--Robertson--Walker--Lema{\v\i}tre spaces are of embbeding class
one (Robertson, 1933). 

\bigskip\bigskip

\centerline{\bf C. Results Related to the Class of the Embedding}

\bigskip

In this section we list several articles containing results specifically
related to the class of the embedding of relevance for general
relativity. Some articles containing results related to the class of the
embedding but other more general results has already been listed in A or
B. They are divided in: class 1; class 2; spherically symmetric fields
(class 2); embedding of the Schwarzschild solution; and spherically
symmetric fields of class 1.

\bigskip\bigskip

\centerline{\bf D. Extrinsic Gravity}

\bigskip

Since the pioneer work of Sakharov in 1966, several attempts have been
done in order to incorporate the concept of bending of space--time in
gravitational theories. One way of doing this is by considering
space--time as embedded in a higher--dimensional space. In this case the
concept of bending has an inmediate geometric visualisation. The first
attempt in this direction seems to be due to Regge and Teitelboim (1976).
The idea was later revived by Pav\v s\i\v c in the eighties, and by Maia
and Tapia in 1989.

\bigskip\bigskip

\centerline{\bf E. Strings and Membranes}

\bigskip

During the 80's there was a good lot of work on string theories. Since
the string is a (1--dimensional) geometrical object living in a
higher--dimensional space the use of embedding should be an inmediate
tool. However, even when the main developments in strings were in other
directions there was also some work dealing with the geometrical,
embedding, aspects of strings. Several proposals were done using
embeddings and generalisations to other higher--dimensional (membranes)
objects were studied.

\bigskip\bigskip

\centerline{\bf F. Rigid Particles and Zitterbewegung}

\bigskip

The Dirac equation gives rise to the phenomenon of zitterbewegung
(Schr\"odinger, 1930). Soon after that, it was realized that this motion
can be explained by a helical motion of the electron (Huang, 1952;
Corben, 1961). Later on, it was discovered that a helical motion can be
described by an acceleration dependent Lagrangian (Liebowitz, 1969;
Riewe, 1971, 1972). The relativistic generalizations making use of
embedding geometry are many and are listed in this section.

\bigskip\bigskip

\centerline{\bf G. New Brane World}

\bigskip

A different approach to considering our universe as embedded in a 
higher--dimensional flat space  was developed mainly by Akama (1978, 1983).
A similar approach was pursued by Pav\v si\v c (1994,1996,1997,1998),
who, in addition, proposed a model in which massive matter on
the brane is due to the intersections with other branes (Pav\v si\v c, 1986).
Recently the idea that our universe is a 3-brane moving in a higher
dimensional space has been revived, but this time as a consequence of 
developments in other areas, mainly in string theory, $M$--theory and 
the like, with the proposal due to Randam and Sundrum (1999). 
The analysis of the Randam and Sundrum model in geometrical terms is 
discussed in (Anchordoqui, 2000).

\bigskip\bigskip

\centerline{\bf Acknowledgements}

\bigskip

This work was partially done during a visit of the authors to The {\bf
Abdus Salam} International Centre for Theoretical Physics, Trieste,
Italy.

\bigskip\bigskip

\centerline{\bf A. The Geometry of Embeddings}

\bigskip

\begin{enumerate}

\item K. F. Gauss, {\it Disquisitiones generales circa superficies
curvas} (1827). English translation: {\it General Investigations of
Curved Surfaces of 1827 and 1825} by C.J. Morehead and A.M. Hiltebeitel
(New York: Raven Press, 1965). In Karl Friedrich Gauss Werke, IV (1827).

\item B. Riemann, {\it \"Uber die Hypothesen welche der Geometrie zu
Grunde liegen} (1854). This thesis was presented on June 10th, 1854, in
G\"ottingen and it was first published in {\it Abh. K\"onigl. gesellsch.}
{\bf 13}, 1 (1868). It was translated into English by W.K. Clifford and
published in {\it Nature} {\bf 8}, 14 (1873). Since the work by Riemann
was published only in 1868, this is the date usually used in the
references in order to avoid confusion over the order of appearance, and
precedence, of ideas in geometry.

\item L. Schl\"afli, {\it Nota alla Memoria del Signor Beltrami, ``Sugli
spaz{\`\i}i di curvatura costante''}, Ann. di Mat. 2$^e$ s\'erie, {\bf
5}, 170 (1871).

\item M. Janet, {\it Sur la possibilit\'e du plonger un espace riemannien
donn\'e dans un espace euclidien}, Ann. Soc. Polon. Math. {\bf 5}, 38
(1926).

\item J. E. Campbell, {\it A Course of Differential Geometry} (Clarendon,
Oxford, 1926).

\item J. Lense, {\it \"Uber ametrische Mannifaltigkeiten und quadratische
Differentialformen mit verchwindender Diskriminante}, Jber. Deutsche
Math. {\bf 35}, 280 (1926).

\item L. P. Eisenhart, {\it Riemannian Geometry} (Princeton University
Press, Princeton, 1926).

\item E. Cartan, {\it Sur la possibilit\'e de plonger un espace
Riemannien donn\'e dans un espace euclidien}, Ann. Soc. Polon. Math. {\bf
6}, 1 (1927).

\item C. Burstin, {\it Ein Beitrag zum Problem der Einbettung der
riemannschen Raume in Euklidischen Raumen}, Rec. Math. Moscou (Math.
Sbornik) {\bf 38}, 74 (1931).

\item T. Y. Thomas, {\it Riemannian spaces of class one and their
characterisation}, Acta Math. {\bf 67}, 169 (1936).

\item J. A. Schouten and D. J. Struik, {\it Einf\"urung in die neueren
Methoden der Differentialgeometrie}, Bd. 2 (Groninggen, 1938), S. 142.

\item T. Y. Thomas, {\it Imbedding theorems in differential geometry},
Bull. Am. Math. Soc. {\bf 45}, 841 (1939).

\item N. A. Rosenson, {\it On Riemannian spaces of class I. Parts I, II,
III}, Izvest. Akad. Nauk SSSR {\bf 4}, 181 (1940); {\bf 5}, 325 (1941);
{\bf 7}, 253 (1943); in Russian.

\item M. Matsumoto, {\it Riemann spaces of class two and their algebraic
characterisation. I, II, III}, J. Math Soc. (Japan) {\bf 2}, 67, 77, 87
(1950).

\item M. Matsumoto, {\it On the special Riemann spaces of class two},
Mem. Coll. Sci. Univ. Kyoto A math. {\bf 26}, 149 (1951).

\item K. Leichtweiss, {\it Das problem von Cauchy in der
mehrdimensionalen Differentialgeometrie I}, Math. Ann. {\bf 130}, 442
(1956).

\item J. Nash, {\it The imbedding problem for Riemannian manifolds}, Ann.
Math. {\bf 63}, 20 (1956).

\item B. O'Neill, {\it An algebraic criterion for inmersion}, Pacific J.
Math. {\bf 9}, 1239 (1959).

\item A. Friedman, {\it Local isometric embedding of Riemannian manifolds
with indefinite metric}, J. Math. Mech. {\bf 10}, 625 (1961).

\item C. J. Clarke, {\it On the isometric global embedding of
pseudo--riemannian manifolds}, Proc. Roy. Soc. London A {\bf 314}, 417
(1970).

\item R. E. Greene, {\it Isometric embedding of Riemannian and
pseudo--Riemannian manifolds}, Memoirs Am. Math. Soc. n. 97 (1970).

\item H. Rund, {\it Invariant Theory of Variational Problems on Subspaces
of a Riemannian Manifold} (G\"ottingen, Vandenhoeck and Ruprecht, 1971).

\item W. B. Bonnor, {\it Null hypersurfaces in Minkowski space--time},
Tensor, N. S. {\bf 24}, 329 (1972).

\item H. Goenner, {\it On the interdependence of the
Gauss--Codazzi--Ricci equations of local isometric embedding}, Gen. Rel.
Grav. {\bf 8}, 139 (1977).

\item M. Gunther, {\it Zum Einbettungssatz von J. Nash}, Math. Nachr.
{\bf 144}, 165 (1989).

\item K. Kobayashi, {\it Fundamental equations for submanifolds},
Fortschr. Phys. {\bf 37}, 599 (1989).

\end{enumerate}

\bigskip\bigskip

\centerline{\bf B. Applications in General Relativity}

\bigskip

\begin{enumerate}

\item A. Einstein, {\it Zur Allgemeinen Relativitatstheorie}, Preuss.
Akad. Wiss. Berlin, Sitzber., 778 (1915).

\item E. Kasner, {\it The impossibility of Einstein fields inmersed in
flat space of five dimensions}, Am. J. Math. {\bf 43}, 126 (1921).

\item E. Kasner, {\it Finite representation of the solar gravitational
field in flat space of six dimensions}, Am. J. Math. {\bf 43}, 130 (1921).

\item J. A. Schouten and D. J. Struik, {\it On some properties of general
manifolds relating to Einstein`s theory of gravitation}, Am. J. Math.
{\bf 43}, 213 (1921)..

\item E. Kasner, {\it Geometrical theorems on Einstein's cosmological
equations}, Am. J. Math. {\bf 43}, 217 (1921).

\item H. P. Robertson, {\it On the foundations of relativistic
cosmology}, Proc. Nat. Acad. Sci. {\bf 15}, 822 (1929).

\item H. P. Robertson, {\it Relativistic cosmology}, Rev. Mod. Phys. {\bf
5}, 62 (1933).

\item A. Gi\~ao, {\it The equations of Codazzi and the relations between
electromagnetism and gravitation}, Phys. Rev. {\bf 76}, 764 (1949).

\item I. Robinson and Y. Ne'eman, {\it Seminar on the Embedding Problem},
Rev. Mod. Phys. {\bf 37}, 201 (1965).

This seminar's report contains the following individual papers:

\bigskip

\item A. Friedman, {\it Isometric embedding of Riemannian manifolds into
Euclidean spaces}, Rev. Mod. Phys. {\bf 37}, 201 (1965).

\item J. Rosen, {\it Embedding of various relativistic Riemannian spaces
in pseudo--Euclidean spaces}, Rev. Mod. Phys. {\bf 37}, 204 (1965).

\item R. Penrose, {\it A remarkable property of plane waves in general
relativity}, Rev. Mod. Phys. {\bf 37}, 215 (1965).

\item C. Fronsdal, {\it Elementary particles in a curved space}, Rev.
Mod. Phys. {\bf 37}, 221 (1965).

\item D. W. Joseph, {\it Generalized covariance}, Rev. Mod. Phys. {\bf
37}, 225 (1965).

\item Y. Ne'eman, {\it Embedded space--time and particle symmetries},
Rev. Mod. Phys. {\bf 37}, 227 (1965).

\bigskip

\item P. Szekeres, {\it Embedding properties of general relativistic
manifolds}, Nuovo Cimento {\bf 43}, 1062 (1966).

\item C. D. Collinson, {\it Embedding of the plane-fronted waves and
other space--times}, J. Math. Phys. {\bf 9}, 403 (1968).

\item P. Rastall, {\it Inmersions of space--time}, Canad. J. Phys. {\bf
47}, 607 (1969).

\item I. Goldman and N. Rosen, {\it A universe embedded in a
five--dimensional flat space}, Gen. Rel. Grav. {\bf 2}, 367 (1971).

\item J. Krause, {\it Embedding approach to a restricted covariance group
in general relativity}, Nuovo Cimento B {\bf 18}, 302 (1973).

\item G. H. Goedecke, {\it On global embedding}, J. Math. Phys. {\bf 15},
789 (1974).

\item G. H. Goenner, {\it Local isometric embedding of Riemannian
manifolds with groups of motion}, Gen. Rel. Grav. {\bf 6}, 75 (1975).

\item M. D. Maia, {\it Elementary systems in general relativity}, J.
Math. Phys. {\bf 16}, 1689 (1975).

\item R. Kerner, {\it Deformations of the embedded Einstein spaces}, Gen.
Rel. Grav. {\bf 9}, 257 (1978).

\item M. D. Maia, {\it Symmetry properties of embedded space--times},
Braz. J. Phys. {\bf 8}, 429 (1978).

\item S. Gallone, {\it An extrinsic aspect of geometrodynamics}, Nuovo
Cimento B {\bf 49}, 149 (1979).

\item H. Goenner, {\it Local isometric embedding of Riemannian manifolds
and Einstein's theory of gravitation}, in General Relativity and
gravitation: one hundred years after the birth of Albert Einstein, ed. by
A. Held (Plenum Press, New York, 1980).

\item D. Kramer, H. Stephani, E. Herlt and M. MacCallum, {\it Exact
Solutions of Einstein's Field Equations} (Cambridge University Press,
Cambridge, England, 1980), chapter 32.

\item M. Ferraris, {\it Algebraic isometric embeddings of exact solutions
of the Einstein equations}, Atti Accad. Sci. Torino Cl. Sci. Fis. Mat.
Natur. {\bf 114}, 229 (1980/81).

\item M. D. Maia, {\it Contacts of space--times}, J. Math. Phys. {\bf
22}, 538 (1981).

\item H. Stephani, {\it Embedding}, in {\it Unified Field Theories of
More than 4 Dimensions Including Exact Solutions}, ed. by V. De Sabbata
and E. Schmutzer (World Scientific, Singapore, 1983).

\item J. Ipser and P. Sikivie, {\it Gravitationally repulsive domain
walls}, Phys. Rev. D {\bf 30}, 712 (1984).

\item M. D. Maia, {\it Combined symmetries in curved space--times}, J.
Math. Phys. {\bf 25}, 2090 (1984).

\item M. D. Maia, {\it The physics of the Gauss--Codazzi--Ricci
equations}, J. Comput. Appl. Math. {\bf 5}, 283 (1986).

\item M. D. Maia, {\it Geometric space--time perturbation. I.
Multiparameter perturbations}, J. Math. Phys. {\bf 28}, 647 (1987).

\item M. D. Maia, {\it Geometric space--time perturbation. II. Gauge
invariance}, J. Math. Phys. {\bf 28}, 651 (1987).

\item I. Ozsv\'ath and L. Sapiro, {\it The Sch\"ucking problem}, J. Math.
Phys. {\bf 28}, 2066 (1987).

\item I. Ozsv\'ath, {\it An embedding problem}, J. Math. Phys. {\bf 29},
825 (1988).

\item M. D. Maia and W. L. Roque, {\it Higher--order space--time contacts
and equivalence}, J. Math. Phys. {\bf 32}, 2722 (1991).

\item I. Ozsv\'ath, {\it Embedding problem. II}, J. Math. Phys. {\bf 33},
229 (1991).

\item G. Abolghasem, A. A. Coley and D. J. McManus, {\it Induced matter
theory and embeddings in Riemann flat space--times}, J. Math. Phys. {\bf
37}, 361 (1996).

\item J. E. Lidsey, C. Romero, R. Tavakol and S. Rippl, {\it On
applications of Campbell's embedding theorem}, Class. Quantum Grav. {\bf
14}, 865 (1997).

\item J. E. Lidsey, R. Tavakol and C. Romero, {\it Campbell's embedding
theorem}, Mod. Phys. Lett. A {\bf 12}, 2319 (1997).

\item S. Deser and O. Levin, {\it Equivalence of Hawking and Unruh
temperatures and entropies through flat space embeddings}, Class. Quantum
Grav. {\bf 15}, L85 (1998).

\item J. E. Lidsey, {\it Embedding of superstring backgrounds in Einstein
gravity}, Phys. Lett. B {\bf 417}, 33 (1998).

\end{enumerate}

\bigskip\bigskip

\centerline{\bf C. Results Related to the Class of the Embedding}

\bigskip

Embedding class 1:

\begin{enumerate}

\item K. P. Singh and S. N. Pandey, {\it Riemannian fourfolds of class
one and gravitation}, Proc. Nat. Inst. Sci. India {\bf 26}, 665 (1960).

\item H. Stephani, {\it \"Uber L\"osungen der Einsteinschen
Feldgleichungen die sich in einen f\"unfdimensionalen flachen Raum
einbetten lassen}, Commun. Math. Phys. {\bf 4}, 137 (1967).

\item C. D. Collinson, {\it Einstein--Maxwell fields of embedding class
one}, Commun. Math. Phys. {\bf 8}, 1 (1968).

\item H. Stephani, {\it Einige L\"osungen der Einsteinschen
Feldgleichungen mit idealer Fl\"ussigkeit die sich in einen
f\"unfdimensionalen Raum einbetten lassen}, Commun. Math. Phys. {\bf 9},
53 (1968).

\item C. F. Jex, {\it A criticism of Kasner's proof that no solution of
$G_{\mu\nu}=0$ can be embedded in a flat five--space}, Nuovo Cimento B
{\bf 55}, 77 (1968).

\item S. N. Pandey and I. D. Kansal, {\it Impossibility of class one
electromagnetic fields}, Proc. Cambridge Phil. Soc. {\bf 66}, 153 (1969).

\item S. N. Pandey and Y. K. Gupta, {\it Space--time of class one and
perfect fluid distribution}, Univ. Roorkee Res. J. {\bf 12}, 9 (1970).

\item C. D. Collinson, {\it A theorem on the local isometric embedding of
empty space--times in a space of constant curvature}, J. Phys. A: Math.
Gen. {\bf 4}, 206 (1971).

\item R. N. Tiwari, {\it Riemannian fourfolds of class one}, Gen. Rel.
Grav. {\bf 3}, 211 (1971).

\item G. P. Pokhariyal, {\it Perfect fluid distribution in class one
space time}, Gen. Rel. Grav. {\bf 3}, 87 (1972).

\item A. Barnes, {\it Space--times of embedding class one in general
relativity}, Gen. Rel. Grav. {\bf 5}, 147 (1974).

\item Y. K. Gupta and S. N. Pandey, {\it Electromagnetic fields of
embedding class one}, Indian J. Pure Appl. Math. {\bf 6}, 1388 (1975).

\item J. Krause, {\it The de Donder coordinate condition and minimal
class--1 space--time}, J. Math. Phys. {\bf 16}, 1090 (1975).

\item H. Goenner, {\it Intrinsically and energetically rigid space--times
of class one}, Tensor (N.S.) {\bf 30}, 15 (1976).

\item J. Krause, {\it Structure of the Einstein tensor for class--1
embedded space--time}, Nuovo Cimento B {\bf 32}, 381 (1976).

\item G. P. Pokhariyal, {\it Perfect fluid distribution in class one
space time. II}, Prog. Math. (Allahabad) {\bf 10}, 11 (1976).

\item S. N. Pandey and S. P. Sharma, {\it Plane symmetric space--time of
class 1}, Gen. Rel. Grav. {\bf 8}, 147 (1977).

\item R. Tikekar, {\it A note on plane symmetric perfect fluid
distributions of embedding class one}, Indian J. Pure Appl. Math. {\bf
11}, 1681 (1980).

\item K. P. Singh and O. N. Singh, {\it A class one cosmological model of
electrified viscous fluid}, Proc. Nat. Acad. Sci. India A {\bf 53}, 135
(1983).

\item Y. K. Gupta and S. P. Sharma, {\it Eisntein--Rosen cylindrically
symmetric space--time subject to class-one conditions}, Gen. Rel. Grav.
{\bf 16}, 349 (1984).

\item Y. K. Gupta and S. P. Sharma, {\it Self--gravitating fluids of
class one with nonvanishing Weyl tensor}, J. Math. Phys. {\bf 25}, 3510
(1984).

\item Y. K. Gupta and R. S. Gupta, {\it Nonstatic analogues of
Kohler--Chao solution of imbedding class one}, Gen. Rel. Grav. {\bf 18},
641 (1986).

\item R. Fuentes V., J. L. Lopez B., G. Ovando Z. and T. Matos, {\it
Space--time of class one}, Gen. Rel. Grav. {\bf 21}, 777 (1989).

\item D. Ladino--Luna and J. L. L\'opez B., {\it Space--time of class
one}, Rev. Mex. Fis. {\bf 35}, 623 (1989); In Spanish.

\item D. Ladino--Luna, {\it  Spacetime of class one with perfect fluid},
Rev. Mex. Fis. {\bf 36}, 177 (1990).

\item D. Ladino--Luna, J. L. L\'opez B., J. Morales R. and G. Ovando Z.,
{\it $R_4$ embedded in $E_5$}, Rev. Mex. Fis. {\bf 36}, 354 (1990); In
Spanish.

\item Abdussattar and D. K. Singh, {\it A plane symmetric Lichnerowicz
universe of class one}, Prog. Math. (Varanasi) {\bf 25}, 121 (1991).

\item S. R. Roy and A. Prasad, {\it An L.R.S. Bianchi type V model of
local embedding class one}, Prog. Math (Varanasi) {\bf 25}, 79 (1994).

\item S. R. Roy and A. Prasad, {\it Some L.R.S. Bianchi type V
cosmological models of local embedding class one}, Gen. Rel. Grav. {\bf
26}, 939 (1994).

\item Y. K. Gupta and J. R. Sharma, {\it Non--static non--conformally
flat fluid plates of embedding class one}, Gen. Rel. Grav. {\bf 28}, 1447
(1996).

\bigskip

Embeddings of Class 2:

\bigskip

\item C. D. Collinson, {\it Empty space--times of embedding class two},
J. Math. Phys. {\bf 7}, 608 (1966).

\item H. J. Efinger, {\it Embedding of a relativistic charged particle},
Commun. Math. Phys. {\bf 2}, 55 (1966).

\item M. Sh. Yakupov, {\it On Einstein spaces of embedding class two},
Dokl. Akad. Nauk SSSR {\bf 180}, 1096 (1968); in Russian.

\item M. S. Jakupov, {\it An algebraic chracterization of Einstein spaces
of class two}, Gravitation and the Theory of Relativity {\bf 4--5}, 78
(1968); In Russian.

\item M. Sh. Yakupov, {\it Einstein spaces of class two}, Grav. i Teor.
Otnos., Univ Kazan {\bf 9}, 109 (1973); In Russian.

\item Y. K. Gupta and P. Goel, {\it Class II analogue of T. Y. Thomas's
theorem and different types of embeddings of static spherically symmetric
space--times}, Gen. Rel. Grav. {\bf 6}, 499 (1975).

\item D. E. Hodgkinson, {\it Empty space--times of embedding class two},
Gen. Rel. Grav. {\bf 16}, 569 (1984).

\item D. E. Hodgkinson, {\it Type $D$ empty space--times of embedding
class 2}, Gen. Rel. Grav. {\bf 19}, 253 (1987).

\item M. R. Meskhishvili, {\it Einstein--Maxwell fields with cosmological
term of embedding class one}, Soobshch. Akad. Nauk Gruzzi {\bf 142}, 297
(1991).

\item N. Van den bergh, {\it Lorentz-- and hyperrotation--invariant
classification of symmetric tensors and the embedding class--2 problem},
Class. Quantum Grav. {\bf 13}, 2817 (1996).

\item N. Van den Bergh, {\it Vacuum solutions of embedding class 2;
Petrov types D and N}, Class. Quantum Grav. {\bf 13}, 2839 (1996).

\item N. Van den Bergh, {\it A classification of embedding class 2
vacua}, Helv. Phys. Acta {\bf 69}, 325 (1996).

\bigskip

Spherically Symmetric Embeddings (Class 2):

\bigskip

\item J. Eiesland, {\it The group of motions of an Einstein space},
Trans. Amer. Math. Soc. {\bf 27}, 213 (1925).

\item J. Pleba\~nski, {\it Spherically symmetric worlds as embedded in a
flat space}, Bull. Acad. Polon. Sci. Ser. Math. Astron. Phys. (10), 373
(1962).

\item M. Ikeda, S. Kitamura and M. Matsumoto, {\it On the embedding of
spherically symmetric space--times}, J. Math. Kyoto Univ. {\bf 3}, 71
(1963).

\item G. Sassi, {\it Parametrical embedding of static spherically
symmetric spacetimes}, Gen. Rel. Grav. 23, 367 (1991).

\bigskip

Embedding of the Schwarzschild and related metrics:

\bigskip

\item C. Fronsdal, {\it Completion and embedding of the Schwarzschild
solution}, Phys. Rev. {\bf 116}, 778 (1959).

\item M. D. Kruskal, {\it Maximal extension of Schwarzschild metric},
Phys. Rev. {\bf 119}, 1743 (1960).

\item T. Fujitani, M. Ikeda and M. Matsumoto, {\it On the embedding of
the Schwarzschild space--time. I, II, III}, J. Math. Kyoto Univ. {\bf 1},
43 (1961); {\bf 1}, 63 (1961); {\bf 2}, 255 (1962).

\item J. Rosen, {\it Embedding of the Schwarzschild and Reissner--Weyl
solutions}, Nuovo Cimento {\bf 38}, 631 (1965).

\item M. Ferraris and M. Francaviglia, {\it An algebraic embedding of
Schwarzschild space--time}, Atti Accad. Sci. Torino Cl. Sci. Fis. Mat.
Natur. {\bf 111}, 315 (1977).

\item M. Ferraris and M. Francaviglia, {\it An algebraic isometric
embedding of Kruskal space--time}, Gen. Rel. Grav. {\bf 10}, 283 (1979).

\item M. Ferraris and M. Francaviglia, {\it Algebraic isometric
embeddings of charged spherically symmetric space--times}, Gen. Rel.
Grav. {\bf 12}, 791 (1980).

\item C. F. Chyba, {\it Time--dependent embeddings for
Schwarzschild--like solutions to the gravitational field equations}, J.
Math. Phys. {\bf 23}, 1662 (1982).

\item M. P. Korkina and S. V. Buts, {\it Embedding of the
Schwarzschild--Friedmann model into a six--dimensional flat space},
Russian Phys. J. {\bf 37}, 59 (1994).

\item J. F. Plebanski, {\it An embedding of a Schwarzschild black hole in
terms of elementary functions}, Acta Phys. Polon. B {\bf 26}, 875 (1995).

\bigskip

Spherically symmetric embeddings of class 1:

\bigskip

\item K. R. Karmarkar, {\it Gravitational metrics of spherical symmetry
and class one}, Proc. Indian Acad. Sci. A {\bf 27}, 56 (1948).

\item M. Kohler and K. L. Chao, {\it Zentralsymmetrische statische
Schwerefelder mit R\"aumen der Klasse I}, Z. Naturforsch. {\bf 20a}, 1537
(1965).

\item S. Kitamura, {\it The imbedding of spherically symmetric space
times in a Riemannian 5-space of constant curvature. I, II}, Tensor
(N.S.) {\bf 16}, 74 (1965); {\bf 17}, 185 (1966).

\item H. Takeno, {\it On the imbedding problem of spherically symmetric
space--times}, in {\it Perspectives in geometry and relativity}, ed. by
B. Hoffmann (Indiana University Press, Bloomington, Indiana, 1966).

\item S. N. Pandey and I. P. Kansal, {\it Spherically symmetric
space--time of class one and electromagnetism}, Proc. Cambridge Phil.
Soc. {\bf 64}, 757 (1968).

\item J. Krishna--Rao, {\it On spherically symmetric perfect fluid
distributions and class one property}, Gen, Rel. Grav. {\bf 2}, 385 (1971).

\item S. Kitamura, {\it On spherically symmetric and plane symmetric
perfect fluid solutions and their embedding of class one}, Tensor N. S.
{\bf 48}, 169 (1989).

\item M. P. Dabrowski, {\it Isometric embedding of the spherically
symmetric Stephani universe: some explicit examples}, J. Math. Phys. {\bf
34}, 1447 (1993).

\bigskip

Embeddings of class 5. The Kerr metric:

\bigskip

\item R. R. Kuzeev, {\it Inmersion class of a Kerr field}, Gravitatsiya i
Teoriya Otnositelnosti, {\bf 16}, 93 (1980); in Russian.

\item R. R. Kuzeev, {\it Imbedding of Kerr's space--time}, Gravitatsiya i
Teoriya Otnositelnosti, {\bf 18}, 75 (1981); in Russian.

\item N. A. Sharp, {\it On embeddings of the Kerr geometry}, Canad. J.
Phys. {\bf 59}, 688 (1981).

\item R. R. Kuzeev, {\it Imbedding of the Kerr--Newman metric},
Gravitatsiya i Teoriya Otnositelnosti, {\bf 21}, 123 (1984); in Russian.

\item J. R. Arteaga, {\it Involution and embedding of metrics}, Lect.
Mat. {\bf 16}, 13 (1995); In Spanish.

\end{enumerate}

\bigskip\bigskip

\centerline{\bf D. Extrinsic Gravity}

\bigskip

\begin{enumerate}
 
\item A. D. Sakharov, {\it Vacuum quantum fluctuations in curved space
and the theory of gravitation}, Dokl. Akad. Nauk SSSR {\bf 177}, 70,
(1967), in Russian; English translation: Sov. Phys. Dokl. {\bf 12}, 1040
(1968). Reprinted in Gen. Rel. Grav. {\bf 32}, 361 (2000).

\item T. Regge and C. Teitelboim, {\it General relativity a la string: a
progress report}, in {\it Proceedings of the First Marcel Grossmann
Meeting}, Trieste, Italy, 1975, ed. by R. Ruffini (North--Holland,
Amsterdam, 1977).

\item S. Deser, F. A. E. Pirani and D. C. Robinson, {\it New embedding
model of general relativity}, Phys. Rev. D {\bf 14}, 3301 (1976).

\item S. Naka and C. Itoi, {\it Induced gravity in higher--dimensional
spacetime and the mass scale of the Kaluza--Klein theory}, Prog. Theor.
Phys. {\bf 70}, 1414 (1983).

\item M. D. Maia and W. Mecklenburg, {\it Aspects of high--dimensional
theories in embedding spaces}, J. Math. Phys. {\bf 25}, 3047 (1984).

\item M. D. Maia, {\it Geometry of Kaluza--Klein theory. I. Basic
setting}, Phys. Rev. D {\bf 31}, 262 (1985).

\item M. D. Maia, {\it Geometry of Kaluza--Klein theory. II. Field
equations}, Phys. Rev. D {\bf 31}, 268 (1985).

\item M. Pav\v si\v c, {\it On the quantisation of gravity by embedding
spacetime in a higher dimensional space}, Class. Quantum Grav. {\bf 2},
869 (1985).

\item M. Pav\v si\v c, {\it Classical theory of a space--time sheet},
Phys. Lett. A {\bf 107}, 66 (1985).

\item M. D. Maia, {\it On Kaluza--Klein relativity}, Gen. Rel. Grav. {\bf
18}, 695 (1986).

\item M. D. Maia, {\it On the integrability conditions for extended
objects}, Class. Quantum Grav. {\bf 6}, 173 (1989).

\item M. D. Maia and W. L. Roque, {\it Classical membrane cosmology},
Phys. Lett. A {\bf 139}, 121 (1989).

\item V. Tapia, {\it Gravitation a la string}, Clas. Quantum Grav. {\bf
6}, L49 (1989).

\item V. Tapia, {\it On the role of extrinsic Riemannian geometry in
gravitation}, Rev. Mex. Fis. {\bf 36}, S164 (1990).

\item V. A. Franke and V. Tapia, {\it The ADM Lagrangian in extrinsic
gravity}, Nuovo Cimento B {\bf 107}, 611 (1992).

\item M. Pav\v si\v c, {\it On the embedding models of gravity and their
prospects}, in Proc. M. Grossmann Meeting 1991 (World Scientific,
Singapore, 1993).

\item M. D. Maia and G. S. Silva, {\it Geometrical constraints on the
cosmological constant}, Phys. Rev. D {\bf 50}, 7233 (1994).

\item E. M. Monte and M. D. Maia, {\it The twisting connection of
space--time}, J. Math. Phys. {\bf 37}, 1972 (1996).

\item C. Romero, R. Tavakol and R. Zalatednikov, {\it The embedding of
general relativity in five dimensions}, Gen. Rel. Grav. {\bf 28}, 365
(1996).

\item E. M. Monte and M. D. Maia, {\it Geometry of the Einstein and
Yang--Mills equations}, Int. J. Theor. Phys. {\bf 36}, 2827 (1997).

\item S. S. Kokarev, {\it GR and multidimensional elasticity theory},
Grav. Cosmol. {\bf 4}, 96 (1998)

\item S. S. Kokarev, {\it Space--time as multidimensional elastic plate},
Nuovo Cimento B {\bf 113}, 1339 (1998)

\item S. S. Kokarev, {\it Space--time as a strongly bent plate}, Nuovo
Cimento B {\bf 114}, 903 (1999).

\end{enumerate}

\bigskip\bigskip

\centerline{\bf E. Strings and Membranes}

\bigskip

\begin{enumerate}

\item P. A. M. Dirac, {\it An extensible model of the electron}, Proc.
Roy. Soc. A {\bf 268}, 7 (1962).

\item B. M. Barbashov and N. A. Chernikov, {\it Solution and quantization
of a nonlinear two--dimensional model for a Born--Infeld type field},
Soviet Phys. JETP {\bf 23}, 861 (1966).

\item Y. Nambu, {\it Duality and Hydrodynamics}, in {\it Copenhagen
Symposium on Strings} (1970).

\item T. Goto, {\it Relativistic quantum mechanics of one--dimensional
mechanical continuum and subsidiary condition of dual resonance model},
Prog. Theor. Phys. {\bf 46}, 1560 (1971).

\item W. Helfrich, {\it }, Z. Naturforsch. C {\bf 28}, 693 (1973).

\item A. O. Barut, {\it Relativistic Composite Systems and Extensible
Models of Fields and Particles}, (1974).

\item P. A. Collins and R. W. Tucker, {\it Classical and quantum
mechanics of free relativistic membranes}, Nucl. Phys. B {\bf 112}, 150
(1976).

\item P. S. Letelier, {\it Classical dynamics of $(n+1)$--dimensional
minimal surfaces}, (1977).

\item A. Aurilia, D. Christodoulou and F. Legovini, {\it A classical
interpretation of the bag model for hadrons}, Phys. Lett. B {\bf 73}, 429
(1978).

\item A. Aurilia and D. Christodoulou, {\it Dynamics of a relativistic
bubble}, Phys. Lett. B {\bf 78}, 589 (1978).

\item B. M. Barbashov and V. V. Nesterenko, {\it Relativistic string
model in a space--time of constant curvature}, Commun. Math. Phys. {\bf
78}, 499 (1981).

\item A. Sugamoto, {\it }, Nucl. Phys. B {\bf 215}, 381 (1981).

\item B. M. Barbashov, V. V. Nesterenko and A. M. Cheryakov, {\it General
solutions of nonlinear equations in the geometric theory of the
relativistic string}, Commun. Math. Phys. {\bf 84}, 471 (1982).

\item P. G. de Gennes and C. Taupin, {\it }, J. Phys. Chem. {\bf 86},
2294 (1982).

\item B. M. Barbashov and V. V. Nesterenko, {\it B\"acklund
transformation for the Liouville equation and gauge conditions in the
theory of a relativistic string}, Theor. Math. Phys. {\bf 56}, 752
(1983).

\item W. Helfrich, {\it Effect of thermal undulations on the rigidity of
fluid membranes and interfaces}, J. Physique {\bf 46}, 1263 (1985).

\item L. Peliti and S. Leibler, {\it }, Phys. Rev. Lett. {\bf 54}, 1690
(1985).


\item T. L. Curtright, G. I. Ghandour, C. B. Thorn and C. K. Zachos, {\it
Trajectories of strings with rigidity}, Phys. Rev. Lett. {\bf 57}, 799
(1986).

\item T. L. Curtright, G. I. Ghandour and C. K. Zachos, {\it Classical
dynamics of strings with rigidity}, Phys. Rev. D {\bf 34}, 3811 (1986).

\item F. David, {\it Rigid surfaces in a space with large
dimensionality}, Europhys. Lett. {\bf 2}, 577 (1986).

\item D. F\"orster, {\it On the scale dependence, due to thermal
fluctuations, of the elastic properties of membranes}, Phys. Lett. A {\bf
114}, 115 (1986).

\item W. Helfrich, {\it Size distributions of vesicles: the role of the
effective rigidity of membranes}, J. Physique {\bf 47}, 321 (1986).

\item A. Kavalov and A. Sedrakyan, {\it }, Phys. Lett. B {\bf 182}, 33
(1986).

\item H. Kleinert, {\it Thermal softening of curvature elasticity in
membranes}, Phys. Lett. A {\bf 114}, 263 (1986).

\item H. Kleinert, {\it Size distribution of spherical vesicles}, Phys.
Lett. A {\bf 116}, 57 (1986).

\item H. Kleinert, {\it The membrane properties of condensing strings},
Phys. Lett. B {\bf 174}, 335 (1986).

\item L. Peliti, {\it }, Physica A {\bf 146}, 269 (1986).

\item A. M. Polyakov, {\it Fine structure of strings}, Nucl. Phys. B {\bf
268}, 406 (1986).


\item F. Alonso and D. Espriu, {\it On the fine structure of strings},
Nucl. Phys. B {\bf 283}, 393 (1987).

\item J. Ambj\o rn, B. Durhuus, J. Fr\"olich and J. Jonsson, {\it
Regularised strings with extrinsic curvature}, Nucl. Phys. B {\bf 290},
480 (1987).

\item A. Aurilia, R. S. Kissack, R. Mann and E. Spalucci, {\it
Relativistic bubble dynamics: From cosmic inflation to hadronic bags},
Phys. Rev. D {\bf 35}, 2961 (1987).

\item E. Braaten, R. D. Pisarski and S.--M. Tse, {\it Static potential
for smooth strings}, Phys. Rev. Lett. {\bf 58}, 93 (1987).

\item E. Braaten and S.--M. Tse, {\it Static potential for smooth strings
in the large--$D$ limit}, Phys. Rev. D {\bf 36}, 3102 (1987).
 
\item E. Braaten and C. K. Zachos, {\it Instability of the static
solution to the closed string with rigidity}, Phys. Rev. D {\bf 35}, 1512
(1987).

\item F. David and E. Guitter, {\it Instabilities in membrane models},
Euorphys. Lett. {\bf 3}, 1169 (1987).

\item D. Foerster, {\it Tangential flows in fluid membranes and their
effect on the softening of curvature rigidity with scale}, Europhys.
Lett. {\bf 4}, 65 (1987).

\item K. Fujikawa and J. Kubo, {\it On the quantization of membrane
theories}, Phys. Lett. B {\bf 199}, 75 (1987).

\item W. Helfrich, {\it Measures of integration in calculating the
effective rigidity of fluid surfaces}, J. Physique {\bf 48}, 285 (1987).

\item J. Hoppe, {\it Relativistic minimal surfaces}, Phys. Lett. B {\bf
196}, 451 (1987).

\item H. Kleinert, {\it }, Phys. Rev. Lett. {\bf 58}, 1915 (1987).

\item P. O. Mazur and V. P. Nair, {\it }, Nucl. Phys. B {\bf 284}, 146
(1987).

\item P. Olesen and S.--K. Yang, {\it }, Nucl. Phys. B {\bf 283}, 73
(1987).

\item M. Pav\v si\v c, {\it Phase space action for minimal surfaces of
any dimension in curved space--time}, Phys. Lett. B {\bf 197}, 327 (1987).

\item R. D. Pisarski, {\it }, Phys. Rev. Lett. {\bf 58}, 1300 (1987).


\item R. Amorim and J. Barcelos--Neto, {\it Effective Lagrangians for
$p$--branes}, Int. J. Mod. Phys. A {\bf 5}, 2667 (1988).

\item A. O. Barut and M. Pav\v si\v c, {\it The spinning minimal surfaces
without the Grassmann variables}, Lett. Math. Phys. {\bf 16}, 333 (1988).

\item F. David and E. Guitter, {\it Rigid random surfaces at large $d$},
Nucl. Phys. B {\bf 295}, 332 (1988).

\item B. P. Dolan and D. H. Tchrakian, {\it New Lagrangians for bosonic
$m$--branes with vanishing cosmological constant}, Phys. Lett. B {\bf
202}, 211 (1988).

\item J. Gamboa and M. Ruiz--Altaba, {\it Functional diffusion equation
for membranes}, Phys. Lett. B 205, 245 (1988).

\item R. Gregory, {\it Effective action for a cosmic string}, Phys. Lett.
B {\bf 206}, 199 (1988).

\item R. Gregory, {\it Cosmic string actions}, (1988).

\item C.--L. Ho and Y. Hosotani, {\it Field theory of geometric
membranes}, Phys. Rev. Lett. {\bf 60}, 885 (1988).

\item C. Itoi, {\it Extrinsic torsion in rigid strings}, Phys. Lett. B
{\bf 211}, 146 91988).

\item C. Itoi and H. Kubota, {\it BRST quantization of the string model
with extrinsic curvature}, Phys. Lett. B {\bf 202}, 381 (1988).

\item H. Kleinert, {\it Dynamical generation of string tension and
stiffness in strings and membranes}, Phys. Lett. B {\bf 211}, 151 (1988).

\item K. I. Maeda and N. Turok, {\it Finite--width corrections to the
Nambu action for the Nielsen--Olensen string}, Phys. Lett. B {\bf 202},
376 (1988).

\item N. Manko\v c--Bor\v stnik and M. Pav\v si\v c, {\it A systematic
examination of five--dimensional Kaluza--Klein theory with sources
consisting of point particles or strings}, Nuovo Cimento A {\bf 99}, 489
(1988).

\item V. V. Nesterenko and Nguen Suan Han, {\it The Hamiltonian formalism
in the model of the relativistic string with rigidity}, Int. J. Mod.
Phys. A {\bf 3}, 2315 (1988).

\item J. A. Nieto, {\it A relativistic 3--dimensional extenden object:
the terron}, Rev. Mex. Fis. {\bf 34}, 597 (1988).

\item M. Pav\v si\v c, {\it Generalisation of the BDHP string action to
membranes of any dimension in curved spacetime}, Class. Quantum Grav.
{\bf 5}, 247 (1988).

\item M. Pav\v si\v c, {\it Classical motion of membranes, strings and
point particles with extrinsic curvature}, Phys. Lett. B {\bf 205}, 231
(1988).

\item K. S. Viswanathan and X. Zhou, {\it $H$--invariance, imbedding
ghost, and string with extrinsic curvature}, Phys. Lett. B {\bf 202}, 217
(1988).


\item S. M. Barr and D. Hochberg, {\it Fine structure of local and axion
strings}, Phys. Rev. D {\bf 39}, 2308 (1989).

\item F. David, {\it Geometry and field theory of random surfaces and
membranes}, in Statistical Mechanics of Membranes and Surfaces, ed. by D.
R. Nelson, T. Piran and S. Weinberg, Winter School for Theoretical
Physics (World Scientific, Singapore, 1989).

\item E. Eizenberg and Y. Ne'eman, {\it Classical Lagrangian and
Hamiltonian formalisms for elementary extendons}, Nuovo Cimento A {\bf
102}, 1183 (1989).

\item D. H. Hartley, M. \"Onder and R. W. Tucker, {\it On the
Einstein--Maxwell equations for a `stiff' membrane}, Class. Quantum Grav.
{\bf 6}, 1301 (1989).

\item C. Itoi and H. Kubota, {\it Gauge invariance based on the extrinsic
geometry in the rigid string}, Z. Phys. C {\bf 44}, 337 (1989).

\item H. Koibuchi and M. Yamada, {\it Statistical mechanics of a rigid
bubble: higher derivative quantum gravity on a spherical random surface
embedded in $E^3$}, Mod. Phys. Lett. A {\bf 4}, 2417 (1989).

\item H. Luckock amd I. Moss, {\it The quantum geometry of random
surfaces and spinning membranes}, Class. Quantum Grav. {\bf 6}, 1993
(1989).

\item K. M. Short, {\it Strings and membranes in curved space}, Class.
Quantum Grav. {\bf 6}, 1351 (1989).

\item V. Tapia, {\it The geometrical meaning of the harmonic gauge for
minimal surfaces}, Nuovo Cimento B {\bf 103}, 435 (1989).


\item G. Gonnella, {\it A comment to ``regularised strings with extrinsic
curvature''}, Phys. Lett. B {\bf 240}, 77 (1990).

\item J. Hoppe, {\it Membranes and integrable systems}, Phys. Lett. B
{\bf 250}, 44 (1990).

\item M. S. Kafkalidis, {\it A geometric approach to the path integral
formalism of $p$--branes}, J. Math. Phys. 31, 2864 (1990).

\item P. Letelier, {\it Nambu bubbles with curvature corrections}, Phys.
Rev. D {\bf 41}, 1333 (1990).

\item M. N. Stoilov and D. Tz. Stoyanov, {\it A model of $p$--branes with
closed--constraint algebra}, J. Phys. A: Math. Gen. {\bf 23}, 5925 (1990).

\item X. Zhou, {\it Static quark potential for a rigid string with
constant density}, Phys. Lett. B {\bf 238}, 205 (1990).


\item R. Beig, {\it Classical geometry of bosonic string dynamics}, Int.
J. Theor. Phys. {\bf 30}, 211 (1991).

\item G. Germ\'an, {\it Some developments in Polyakov--Kleinert string
with extrinsic curvature stiffness}, Mod. Phys. Lett. A {\bf 6}, 1815
(1991).

\item T. H. Hansson, J. Isberg, U. Lindstr\"om, H. Nordstr\"om and J.
Grundberg, {\it Rigid strings from field theory}, Phys. Lett. B {\bf
261}, 379 (1991).

\item J. A. Nieto and C. Nu\~nez, {\it Strings from Weyl invariant
membranes}, Nuovo Cimento B {\bf 106}, 1045 (1991).

\item T. Onogi, {\it Phase transitions in the Nambu--Goto string model
with extrinsic curvature}, Phys. Lett. B {\bf 255}, 209 (1991).

\item G. M. Sotkov, M. Stanishkov and C.-J. Zhu, {\it Extrinsic geometry
of strings and $W$--gravities}, Nucl. Phys. B {\bf 356}, 245 (1991).

\item P. W\c egrzyn, {\it Edge conditions for the open string with
rigidity}, Phys. Lett. B {\bf 269}, 311 (1991).


\item A. O. Barut and M. Pav\v si\v c, {\it A covariant canonical
decomposition of the bosonic and spinning extended objects with electric
charge}, Mod. Phys. Lett. A {\bf 7}, 1381 (1992).

\item B. Carter, {\it Basic brane theory}, Class. Quantum Grav. {\bf 9}
S19 (1992).

\item V. V. Nesterenko and N. R. Shvetz, {\it The Casimir energy of the
rigid string with massive ends}, Z. Phys. C {\bf 55}, 265 (1992).

\item M. Pav\v si\v c, {\it Point particle--like action for $p$--branes},
Class. Quantum Grav. {\bf 9}, L13 (1992).


\item A. Antill\'on and G. Germ\'an, {\it Numerical study of the
Nambu--Goto string model at finite lenght and temperature}, Phys. Rev. D
{\bf 47}, 4567 (1993).

\item A. Aurilia and E. Spallucci, {\it Gauge theory of relativistic
membranes}, Class. Quantum Grav. {\bf 10}, 1217 (1993).

\item A. O. Barut and M. Pav\v si\v c, {\it Dirac's shell model of the
electron and the general theory of moving relativistic chraged
membranes}, Phys. Lett. B {\bf 306}, 49 (1993).

\item B. Carter, {\it Perturbation dynamics for membranes and strings
governed by Dirac Goto Nambu action in curved space}, Phys. Rev. D {\bf
48}, 4853 (1993).

\item A. It\^o, {\it Skyrme--like Lagrangian and Nambu--Goto--type
action}, Nuovo Cimento A {\bf 106}, 423 (1993).

\item V. Silveira and M. D. Maia, {\it Topological defects and
corrections to the Nambu action}, Phys. Lett. A {\bf 174}, 280 (1993).


\item A. O. Barut and M. Pav\v si\v c, {\it Radiation reaction and the
electromagnetic energy--momentum tensor of moving relativistic charged
membranes}, Phys. Lett. B {\bf 331}, 45 (1994).

\item B. Carter, {\it Equations of motion of a stiff geodynamic string or
higher brane}, Class. Quantum Grav. {\bf 11}, 2677 (1994).

\item P. Orland, {\it Extrinsic curvature dependence of Nielsen--Olesen
strings}, Nucl. Phys. B {\bf 428}, 221 (1994).

\item J. F. Wheater, {\it Random surfaces: from polymer membranes to
strings}, J. Phys. A: Math. Gen. {\bf 27}, 3323 (1994).


\item R. A. Battye and B. Carter, {\it Gravitational perturbations of
relativistic membranes and strings}, Phys. Lett. B {\bf 357}, 29 (1995).

\item R. Capovilla and J. Guven, {\it Algebra of surface deformations},
Class. Quantum Grav. {\bf 12}, L107 (1995).

\item R. Capovilla and J. Guven, {\it Geometry of deformations of
relativistic membranes}, Phys. Rev. D {\bf 51}, 6736 (1995).

\item R. Capovilla and J. Guven, {\it Large deformations of relativistic
membranes: a generalization of the Raychaudhuri equations}, Phys. Rev. D
{\bf 52}, 1072 (1995).

\item A. L. Kholodenko  and V. V. Nesterenko, {\it Classical dynamics of
rigid string from the Willmore functional}, J. Geom. Phys. {\bf 16}, 15
(1995).

\item Y. Ne'eman and E. Eizenberg, {\it Membranes and other extendons:
p--branes} (World Scientific, Singapore, 1995).

\item M. Pav\v si\v c, {\it Relativistic $p$--branes without constraints
and their relation to the wiggly extended objects}, Found. Phys. {\bf
25}, 819 (1995).

\item M. Pav\v si\v c, {\it The classical and quantum theory of
relativistic $p$--branes without constraints}, Nuovo Cimento A {\bf 108},
221 (1995).


\item I. A. Bandos, {\it On a zero curvature representation for bosonic
strings and $p$ branes}, Phys. Lett. B {\bf 388}, 35 (1996).

\item R. Capovilla, R. Cordero and J. Guven, {\it Conformal invariants of
the extrinsic geometry of relativistic membranes}, Mod. Phys. Lett. A
{\bf 11}, 2755 (1996).

\item S. Kar, {\it Generalized Raychaudhuri equations: Examples}, Phys.
Rev. D {\bf 53}, 2071 (1996).

\item S. Kar, {\it Generalized Raychaudhuri equations for strings in the
presence of an antisymmetric tensor field}, Phys. Rev. D {\bf 54}, 6408
(1996).

\item A. L. Larsen and C. O. Lousto, {\it On the stability of spherical
membranes in curved space--time}, Nucl. Phys. B {\bf 472}, 361 (1996).

\item M. Pav\v si\v c, {\it The relativistic charged membrane and its
total mass}, Helv. Phys. Acta {\bf 69}, 353 (1996).

\item K. J. Wiese, {\it Classification of perturbations for membranes
with bending rigidity}, Phys. Lett. B {\bf 387}, 57 (1996).


\item M. Abou Zeid and C. M. Hull, {\it Intrinsic geometry of
$D$--branes}, Phys. Lett. B {\bf 404}, 264 (1997).

\item R. Capovilla and J. Guven, {\it Extended objects with edges}, Phys.
Rev. D {\bf 55}, 2388 (1997).

\item R. Capovilla and J. Guven, {\it Geometry of composite relativistic
extended objects}, Nucl. Phys. B (Proc. Suppl.) {\bf 57}, 273 (1997).

\item V. V. Nesterenko and I. G. Pirozhenko, {\it Justification of the
zeta function renormalization in rigid string model}, J. Math. Phys. {\bf
38}, 6265 (1997).

\item M. Pav\v si\v c, {\it Fock--Schwinger proper time formalism for
$p$--branes}, Nucl. Phys. Proc. Suppl. {\bf 57}, 265 (1997).

\item M. Pav\v si\v c, {\it The Dirac--Nambu--Goto $p$--branes as
particular solutions to a generalized unconstrained system}, Nuovo
Cimento A {\bf 110}, 369 (1997).


\item M. Abou Zeid and C. M. Hull, {\it Geometric actions for $D$--branes
and $M$--branes}, Phys. Lett. B {\bf 428}, 277 (1998).

\item C. G. Callan and J. M. Maldacena, {\it Brane dynamics from the
Born--Infeld action}, Nucl. Phys. B {\bf 513}, 198 (1998).

\item R. Capovilla and J. Guven, {\it Open strings with topologically
inspired boundary conditions}, Class. Quantum Grav. {\bf 15}, 1111 (1998).

\item R. Capovilla and J. Guven, {\it Deformations of extended objects
with edges}, Phys. Rev. D {\bf 57}, 5158 (1998).

\item G. W. Gibbons, {\it Born--Infeld particles and Dirichlet
$p$--branes}, Nucl. Phys. B {\bf 514}, 603 (1998).

\item V. V. Nesterenko and I. G. Pirozhenko, {\it Open rigid string with
Gauss--Bonnet term in the action}, Mod. Phys. Lett. A {\bf 13}, 2513
(1998).

\item M. Pav\v si\v c, {\it The unconstrained Stueckelberger--like
membranes's action as a reduced Dirac--Nambu--Goto action}, Phys. Lett. A
{\bf 242}, 187 (1998).

\item M. Pav\v si\v c, {\it Formulation of a relativistic theory without
constraints}, Found. Phys. {\bf 28}, 1443 (1998).

\item E. Zafiris, {\it Kinematical approach to brane world sheet
deformations in space--time}, Ann. Phys. {\bf 264}, 75 (1998).

\item E. Zafiris, {\it Covariant generalization of Codazzi--Raychaudhuri
and area change equations for relativistic branes}, J. Geom. Phys. {\bf
28}, 271 (1998).

\item E. Zafiris, {\it Generalized Raychaudhuri and area change equations
for classical brane models}, Phys. Rev. D {\bf 58}, 043509 (1998).


\item I. Bandos, E. Ivanov, A. A. Kapustnikov and S. A. Ulanov, {\it
General solution of string inspired nonlinear equations}, J. Math. Phys.
{\bf 40}, 5203 (1999).

\item G. W. Gibbons, {\it Branes as BIons}, Class. Quantum Grav. {\bf
16}, 1471 (1999).


\item G. Arreaga, R. Capovilla and J. Guven, {\it Noether currents for
bosonic branes}, Ann. Phys. {\bf 279}, 126 (2000).

\item R. A. Battye and B. Carter, {\it Second order Lagrangian and
symplectic current for gravitationally perturbed Dirac--Goto--Nambu
strings and branes}, Class. Quantum Grav. {\bf 17}, 3325 (2000).

\item R. Capovilla, J. Guven and E. Rojas, {\it ADM worldvolume
geometry}, Nucl. Phys. B (Proc. Suppl.) {\bf 88}, 337 (2000).

\item J. C. Mart{\'\i}nez, {\it Bending rigidity of closed membranes in a
thermal bath}, Phys. Lett. A {\bf 266}, 260 (2000).

\end{enumerate}

\bigskip\bigskip

\centerline{\bf F. Rigid Particles and Zitterbewegung}

\begin{enumerate}

\item E. Schr\"odinger, Sitzungb. Preuss. Akad. Wiss. Phys.--Math. Kl.
{\bf 24}, 418 (1930).

\item K. Huang, {\it On the zitterbewegung of the Dirac electron}, Am. J.
Phys. {\bf 20}, 479 (1952).

\item B. Liebowitz, {\it A model of the electron}, Nuovo Cimento A {\bf
63}, 1234 (1969).

\item F. Riewe, {\it Generalized mechanics of a spinning particle}, Lett.
Nuovo Cimento {\bf 1}, 807 (1971).

\item F. Riewe, {\it Relativistic classical spinning--particle
mechanics}, Nuovo Cimento B {\bf 8}, 271 (1972).

\item R. D. Pisarski, {\it Theory of curved paths}, Phys. Rev. D {\bf
34}, 670 (1986).

\item M. S. Plyushchay, {\it Canonical quantization and mass spectrum of
relativistic particle analogue of relativistic string with rigidity},
Mod. Phys. Lett. A {\bf 3}, 1299 (1988).

\item H. Arodz, A. Sitarz and P. Wegrzyn, {\it On relativistic point
particles with curvature dependent actions}, Acta Phys. Polon. B {\bf
20}, 921 (1989).

\item V.V. Nesterenko, {\it Singular Lagrangians with higher
derivatives}, J. Phys. A {\bf 22}, 1673 (1989).

\item M. Pav\v si\v c, {\it The quantization of a point particle with
extrinsic curvature leads to the Dirac equation}, Phys. Lett. B {\bf
221}, 264 (1989).

\item M. S. Plyushchay, {\it Massless point particle with rigidity}, Mod.
Phys. Lett. A {\bf 4}, 837 (1989).

\item M. S. Plyushchay, {\it Massive relativistic point particle with
rigidity}, Int. J. Mod. Phys. A {\bf 4}, 3851 (1989).

\item M. Huq, P. I. Obiakor and S. Singh, {\it Point particle with
extrinsic curvature}, Int. J. Mod. Phys. A {\bf 5}, 4301 (1990).

\item J. Isberg, U. Lindstr\"om and H. Nordstr\"om, {\it Canonical
quantisation of a rigid particle}, Mod. Phys. Lett. A {\bf 5}, 2491
(1990).

\item M. Pav\v si\v c, {\it On the consistent derivation of rigid
particles from strings}, Class. Quantum Grav. {\bf 7}, L187 (1990).

\item M. S. Plyushchay, {\it Relativistic massive particle with higher
curvatures as a model for the description of bosons and fermions}, Phys.
Lett. B {\bf 235}, 47 (1990).

\item M. S. Plyushchay, {\it Relativistic zitterbewegung: the model of
spinning particles without Grassmann variables}, Phys. Lett. B {\bf 236},
291 (1990).

\item M. S. Plyushchay, {\it Massless particle with rigidity as a model
for the description of bosons and fermions}, Phys. Lett. B {\bf 243}, 383
(1990).

\item G. Fiorentini, M. Gasperini and G. Scarpetta, {\it Is the electron
a ``rigid'' particle?}, Mod. Phys. Lett. A {\bf 6}, 2033 (1991).

\item D. G. C. McKeon, {\it Classical motion of a point particle with
extrinsic curvature}, Can. J. Phys. {\bf 69}, 830 (1991).

\item V. V. Nesterenko, {\it Canonical quantization of a relativistic
particle with torsion}, Mod. Phys. Lett. A {\bf 6}, 719 (1991).

\item V. V. Nesterenko, {\it Relativistic particle with curvature in an
external electromagnetic field}, Int. J. Mod. Phys. A {\bf 6}, 3989
(1991).

\item V. V. Nesterenko, {\it Curvature and torsion of the world curve in
the action of the relativistic particle}, J. Math. Phys. {\bf 32}, 3315
(1991).

\item V. V. Nesterenko, {\it Relativistic particle with action that
depends on the torsion of the world trajectory}, Theor. Math. Phys. {\bf
86}, 169 (1991).

\item M. S. Plyushchay, {\it The model of the relativistic particle with
torsion}, Nucl. Phys. B {\bf 362}, 54 (1991).

\item M. S. Plyushchay, {\it Does the quantisation of a particle with
curvature lead to the Dirac equation?}, Phys. Lett. B {\bf 253}, 50 (1991).

\item M. S. Plyushchay, {\it Relativistic particle with torsion, Majorana
equation and fractional spin}, Phys. Lett. B {\bf 262}, 71 (1991).

\item Yu. A. Kuznetsov and M. S. Plyushchay, {\it Tachyonless models of
relativistic particles with curvature and torsion}, Phys. Lett. B {\bf
297}, 49 (1992).

\item V. V. Nesterenko, {\it Torsion in the action of the relativistic
particle}, Class. Quantum Grav. {\bf 9}, 1101 (1992).

\item J. Govaerts, {\it Relativistic rigid particles: classical tachyons
and quantum anomalies},  Z. Phys. C {\bf 57}, 59 (1993).

\item Yu. A. Kuznetsov and M. S. Plyushchay, {\it The model of the
relativistic particle with curvature and torsion}, {\bf 389}, 181 (1993).

\item V. V. Nesterenko, {\it On a model of a relativistic particle with
curvature and torsion}, J. Math. Phys. {\bf 34}, 5589 (1993).

\item Yu. A. Kuznetsov and M. S. Plyushchay, {\it $(2+1)$--dimensional
models of relativistic particles with curbature and torsion}, J. Math.
Phys. {\bf 35}, 2772 (1994).

\item V. V. Nesterenko, A. Feoli and G. Scarpetta, {\it Dynamics of
relativistic particle with Lagrangian dependent on acceleration}, J.
Math. Phys. {\bf 36}, 5552 (1995).

\item M. S. Plyushchay, {\it Relativistic particle with torsion and
charged particle in a constant electromagnetic field: identity of
evolution}, Mod. Phys. Lett. A {\bf 10}, 1463 (1995).

\item B. P. Kosyakov and V. V. Nesterenko, {\it Stability of
zitterbewegung of a rigid particle}, Phys. Lett. B {\bf 384}, 70 (1996).

\item V. V. Nesterenko, A. Feoli and G. Scarpetta, {\it Complete
integrability for Lagrangians dependent on acceleration in a spacetime of
constant curvature}, Class. Quantum Grav. {\bf 13}, 1201 (1996).

\end{enumerate}

\bigskip\bigskip

\centerline{\bf G. New Brane World}

\bigskip

\begin{enumerate}

\item K. Akama, {\it An attempt at pregeometry}, Prog. Theor. Phys. {\bf
60}, 1900 (1978).

\item K. Akama, {\it Pregeometry}, in {\it Proceedings of the Symposium
on Gauge Theory and Gravitation}, Nara, Japan, 1982, Lecture Notes in
Physics 176, ed. by Kikkawa, N. Nakanishi and H. Nariai (Springer,
Berlin, 1982).

\item K. Akama and H. Terazawa, {\it Pregeometric origin of the big
bang}, Gen. Rel. Grav. {\bf 15}, 201 (1983).

\item V. A. Rubakov and M. E. Shaposnikov, {\it Do we live inside a
domain wall?}, Phys. Lett. B {\bf 125}, 136 (1983).

\item V. A. Rubakov and M. E. Shaposnikov, {\it Extra space--time
dimensions: towards a solution to the cosmological constant problem},
Phys. Lett. B {\bf 125}, 139 (1983).

\item M. Visser, {\it An exotic class of Kaluza-Klein models}, Phys.
Lett. B {\bf 159}, 22 (1985).

\item E. J. Squires, {\it Dimensional reduction caused by a cosmological
constant}, Phys. Lett. B {\bf 167}, 286 (1986).

\item M. Pav\v si\v c, {\it String model for general relativity and
canonical formalism for minimal surfaces}, Nuovo Cimento A {\bf 95}, 297
(1986).

\item M. Pav\v si\v c, {\it Einstein's gravity from a first order
Lagrangian in an embedding space}, Phys. Lett. A {\bf 116}, 1 (1986).

\item G. W. Gibbons and D. L. Wiltshire, {\it Spacetime as a membrane in
higher dimensions}, Nucl. Phys. B {\bf 287}, 717 (1987).

\item K. Akama and H. Terazawa, {\it Pregauge--pregeometry and gauge
invariant string model}, Prog. Theor. Phys. {\bf 79}, 740 (1988).

\item K. Akama, {\it Pregeometry and an extended string action}, Progr.
Theor. Phys. {\bf 79}, 1299 (1988).

\item T. Hori, {\it The universe as a relativistic ball}, Phys. Lett. B
{\bf 222}, 188 (1989).

\item R. Floreanini and R. Percacci, {\it Topological pregeometry}, Mod.
Phys. Lett. A {\bf 5}, 2247 (1990)

\item H. Terazawa, {\it Various actions for pregeometry}, Prog. Theor.
Phys. {\bf 86}, 337 (1991).

\item A. Aurilia, A. Smailagic and E. Spallucci, {\it Membrane
pregeometry and the vanishing of the cosmological constant}, Class.
Quantum Grav. {\bf 9}, 1883 (1992).

\item A. A. Bytenko and S. D. Odintsov, {\it The Casimir energy for
bosonic $p$--branes compactified on constant curvature space}, Class.
Quantum Grav. 9, 391 (1992).

\item K. Akama and I. Oda, {\it Towards BRST quantization of pregeometry
and topological pregeometry}, Nucl. Phys. B {\bf 397}, 727 (1993).

\item M. Pav\v si\v c, {\it The embedding model of induced gravity with
bosonic sources}, Found. Phys. {\bf 24}, 1495 (1994).

\item M. Pav\v si\v c, {\it The embedding model of induced gravity with
bosonic sources}, Grav. Cosmol. {\bf 2}, 1 (1996).

\item M. Pav\v si\v c, {\it On the resolution of time problem in quantum
gravity induced from unconstrained membranes}, Found. Phys. {\bf 26}, 159
(1996).

\item I. Bandos, {\it String--like description of gravity and possible
applications for $F$ theory}, Mod. Phys. Lett. A {\bf 12}, 799 (1997).

\item M. Pav\v si\v c, {\it Space--time as an unconstrained membrane},
Grav. Cosmol. {\bf 3}, 305 (1997).


\item A. Davidson and D. Karasik, {\it Quantum gravity of a brane--like
universe}, Mod. Phys. Lett. A {\bf 13}, 2187 (1998).

\item M. Pav\v si\v c, {\it Quantum gravity induced from unconstrained
membranes}, Found. Phys. {\bf 28}, 1465 (1998).


\item I. A. Bandos and W. Kummer, {\it $p$--branes, Poisson sigma models
and embedding approach to $(p+1)$--dimensional gravity}, Int. J. Mod.
Phys. A {\bf 14}, 4881 (1999).

\item A. Davidson, {\it $\Lambda=0$ cosmology of a brane--like universe},
Class. Quantum Grav. {\bf 16}, 653 (1999).

\item A. Davidson, D. Karasik and Y. Lederer, {\it Wavefunction of a
brane--like universe}, Class. Quantum Grav. {\bf 16}, 1349 (1999).

\item W. D. Goldberger, M. B. Wise, {\it Bulk fields in the
Randall--Sundrum compactification scenario}, Phys. Rev. D {\bf 60},
107505 (1999).

\item H. Hatanaka, M. Sakamoto, M. Tachibana and K. Takenaga, {\it Many
brane extension of the Randall--Sundrum solution}, Prog. Theor. Phys.
{\bf 102}, 1213 (1999).

\item N. Kaloper, {\it Bent domain walls as brane--worlds}, Phys. Rev. D
{\bf 60}, 123506 (1999).

\item L. Randall and R. Sundrum, {\it Large mass hierarchy from a small
extra dimension}, Phys. Rev. Lett. {\bf 83}, 3370 (1999).

\item L. Randall and R. Sundrum, {\it An alternative to
compactification}, Phys. Rev. Lett. {\bf 83}, 4690 (1999).


\item L. Anchordoqui and S. P\'erez Bergliaffa, {\it Wormhole surgery and
cosmology on the brane: The world is not enough}, Phys. Rev. D {\bf 62},
067502 (2000).

\item I. Ya. Aref'eva, M. G. Ivanov, W. M\"uck, K. S. Viswanathan and I.
V. Volovich, {\it Consistent linearized gravity in brane backgrounds},
Nucl. Phys. B {\bf 590}, 273 (2000).

\item C. Barcel\'o and M. Visser, {\it Brane surgery: energy conditions,
traversable wormholes, and voids}, Nucl. Phys. B {\bf 584}, 415 (2000).

\item P. Bin\'etruy, J. M. Cline and C. Grojean, {\it Dynamical
instability of brane solutions with a self--tuning cosmological
constant}, Phys. Lett. B {\bf 489}, 403 (2000).

\item P. Bin\'etruy, C. Deffayet, U. Ellwanger and D. Langlois, {\it
Brane cosmological evolution in a bulk with cosmological constant}, Phys.
Lett. B {\bf 477}, 285 (2000).

\item P. Bin\'etruy, C. Deffayet and D. Langlois, {\it Non--conventional
cosmology from a brane universe}, Nucl. Phys. B {\bf 565}, 269 (2000).

\item C. van de Bruck, M. Dorca, R. H. Brandenberger and A. Lukas, {\it
Cosmological perturbations in brane world theories: Formalism}, Phys.
Rev. D {\bf 62}, 123515 (2000).

\item C. van de Bruck, M. Dorca, C. J. A. P. Martins and M. Parry, {\it
Cosmological consequences of the brane/bulk interaction}, Phys. Lett. B
{\bf 495}, 183 (2000).

\item A. Chamblin, S. W. Hawking and H. S. Reall, {\it Brane world black
holes}, Phys. Rev. D {\bf 61}, 065007 (2000).

\item S. Chang, J. Hisano, H. Nakano, N. Okada and M. Yagamuchi, {\it
Bulk standard model in the Randall--Sundrum background}, Phys. Rev. D
{\bf 62}, 084025 (2000).

\item C. Charmousis, R. Gregory and V. A. Rubakov, {\it Wave function of
the radion in a brane world}, Phys. Rev. D {\bf 62}, 067505 (2000).

\item C. Cz\'aki, J. Erlich and T. J. Hollowood, {\it Graviton
propagators, brane bending and bending of light in theories with
quasi--localized gravity}, Phys. Lett. B {\bf 481}, 107 (2000).

\item C. Cs\'aki, J. Erlich, T. J. Hollowood and Y. Shirman, {\it
Universal aspects of gravity localized on thick branes}, Nucl. Phys. B
{\bf 581}, 309 (2000).

\item C. Cs\'aki and Y. Shirman, {\it Brane junctions in the
Randall--Sundrum scenario}, Phys. Rev. D {\bf 61}, 024008 (2000).

\item C. Cs\'aki, M. Graesser, L. Randall and J. Terning, {\it Cosmology
of brane models with radion stabilization}, Phys. Rev. D {\bf 62}, 045015
(2000).

\item N. Dadhich, {\it Negative energy condition and black holes on the
brane}, Phys. Lett. B {\bf 492}, 357 (2000).

\item N. Dadhich, R. Maartens, P. Papadopoulos and V. Rezania, {\it Black
holes on the brane}, Phys. Lett. B {\bf 487}, 1 (2000).

\item S. R. Das, S. P. Trivedi and S. vaidya, {\it Magnetic moments of
branes and giant gravitons}, JHEP 10, 037 (2000).

\item R. Dick, {\it Which action for brane worlds?}, Phys. Lett. B {\bf
491}, 333 (2000).

\item R. Dick and  D. Mikulovi\'c, {\it Gravity and the Newtonian limit
in the Randall--Sundrum model}, Phys. Lett. B {\bf 476}, 363 (2000).

\item S. L. Dubovsky and V. A. Rubakov, {\it Brane world: Disappearing
massive matter}, Phys. Rev. D {\bf 62}, 105011 (2000).

\item S. L. Dubovsky, V. A. Rubakov and P. G. Tinyakov, {\it Is the
electric charge conserved in brane world?}, JHEP 08, 041 (2000).

\item G. Dvali, G. Gabadadze and M. Porrati, {\it A comment on brane
bending and ghosts in theories with infinite extra dimensions}, Phys.
Lett. B {\bf 484}, 129 (2000).

\item G. Dvali, G. Gabadadze and M. Porrati, {\it 4D gravity on a brane
in 5D Minkowski space}, Phys. Lett. B {\bf 485}, 208 (2000).

\item U. Ellwanger, {\it Constraints on a brane world from the vanishing
of the cosmological constant}, Phys. Lett. B {\bf 473}, 233 (2000).

\item E. E. Flanagan, S. H. Henry Tye and I. Wasserman, {\it A cosmology
of the brane world}, Phys. Rev. D {\bf 62}, 024011 (2000).

\item E. E. Flanagan, S. H. Henry Tye and I. Wasserman, {\it Cosmological
expansion in the Randall--Sundrum brane world scenario}, Phys. Rev. D
{\bf 62}, 044039 (2000).

\item S. F\"orste, Z. Lalak, S. Lavignac and H. P. Nilles, {\it A comment
on selftuning and vanishing cosmological constant in the brane world},
Phys. Lett. B {\bf 481}, 360 (2000).

\item S. F\"orste, Z. Lalak, S. Lavignac and H. P. Nilles, {\it The
cosmological constant problem from a brane--world perspective}, JHEP 09,
034 (2000).

\item J. Garriga and M. Sasaki, {\it Brane world creation and black
holes}, Phys. Rev. D {\bf 62}, 043523 (2000).

\item J. Garriga and T. Tanaka, {\it Gravity in the brane world}, Phys.
Rev. Lett {\bf 84}, 2778 (2000).

\item J. Gauntlett, {\it Brane new worlds}, Nature {\bf 404}, 28 (2000).

\item G. Gibbons, {\it Brane--worlds}, Science {\bf 287}, 49 (2000).

\item S. B. Giddings, E. Katz and L. Randall, {\it Linearized gravity in
brane backgrounds}, JHEP 03, 023 (2000).

\item M. Gogberashvili, {\it Our world as an expanding shell},  Europhys.
Lett. {\bf 49}, 396 (2000).

\item M. Gogberashvili, {\it Brane--universe in six dimensions with two
times}, Phys. Lett. B {\bf 484}, 124 (2000).

\item C. G\'omez, B. Janssen and P. Silva, {\it Dilatonic
Randall--Sundrum theory and renormalization group}, JHEP 04, 024 (2000).

\item C. G\'omez, B. Janssen and P. J. Silva, {\it Brane world with bulk
horizons}, JHEP 04, 027 (2000).

\item P. F. Gonz\'alez--D{\'\i}az, {\it Quintessence in brane cosmology},
Phys. Lett. B {\bf 481}, 353 (2000).

\item A. Gorsky and K. Selivanov, {\it Tunneling into the
Randall--Sundrum brane world}, Phys. Lett. B {\bf 485}, 271 (2000).

\item R. Gregory, V. A. Rubakov and S. M. Sibiryakov, {\it Brane worlds"
the gravity of escaping matter}, Class. Quantum Grav. {\bf 17}, 4437
(2000).

\item R. Gregory, V. A. Rubakov and S. M. Sibiryakov, {\it Gravity and
antigravity in a brane world with metastable gravitons}, Phys. Lett. B
{\bf 489}, 203 (2000).

\item S. W. Hawking, T. Hertog and H. S. Reall, {\it Brane new world},
Phys. Rev. D {\bf 62}, 043501 (2000).

\item J. Hisano and N. Okada, {\it On effective theory of brane world
with small tension}, Phys. Rev. D {\bf 61}, 106003 (2000).

\item D. Ida, {\it Brane world cosmology}, JHEP 09, 014 (2000).

\item J. Jiang, T. Li and D. Margatia, {\it Brane networks in AdS space},
Phys. Lett. B {\bf 492}, 187 (2000).

\item Z. Kakushadze, {\it Why the cosmological problem is hard}, Phys.
Lett. B {\bf 488}, 402 (2000).

\item Z. Kakushadze, {\it Bulk supersymmetry and brane cosmology}, Phys.
Lett. B {\bf 489}, 207 (2000).

\item Z. Kakushadze, {\it Conformal brane world and cosmological
constant}, Phys. Lett. B {\bf 491}, 317 (2000).

\item G. Kang and Y. S. Myung, {\it No ghost state in the brane world},
Phys. Lett. B {\bf 483}, 235 (2000).

\item H. B. Kim, {\it Cosmology of Randall--Sundrum models with an extra
dimension stabilized by balancing bulk matter}, Phys. Lett. B {\bf 478},
285 (2000).

\item J. E. Kim and B. Kyae, {\it Exact cosmological solution and modulus
stabilization in the Randall--Sundrum model with bulk matter}, Phys.
Lett. B {\bf 486}, 165 (2000).

\item J. E. Kim, B. Kyae and H. M. Lee, {\it Various modified solutions
of the Randall--Sundrum model with the Gauss--Bonnet interaction}, Nucl.
Phys. B {\bf 582}, 296 (2000); Erratum, ibid {\bf 591}, 587 (2000).

\item J. E. Kim, B. Kyae and H. M. Lee, {\it Effective Gauss--Bonnet
interaction in Randall--Sundrum compactification}, Phys. Rev. D {\bf 62},
045013 (2000).

\item H. Kodama, A. Ishibashi and O. Seto, {\it Brane world cosmology:
Gauge invariant formalism for perturbation}, Phys. Rev. D {\bf 62},
064022 (2000).

\item I. I. Kogan and G. G. Ross, {\it Brane universe and multigravity:
modification of gravity at large and small distances}, Phys. Lett. B {\bf
485}, 255 (2000).

\item K. Koyama and J. Soda, {\it Birth of the brane world}, Phys. Lett.
B {\bf 483}, 432 (2000).

\item D. Langlois, {\it Brane cosmological perturbations}, Phys. Rev. D
{\bf 62}, 126012 (2000).

\item D. Langlois, R. Maartens and D. Wands, {\it Gravitational waves
from inflation on the brane}, Phys. Lett. B {\bf 489}, 259 (2000).

\item J. Lesgourges, S. Pastor, M. Peloso and L. Sorbo, {\it Cosmology of
the Randall--Sundrum model after dilaton stabilization}, Phys. Lett. B
{\bf 489}, 411 (2000).

\item I. Low and A. Zee, {\it Naked singularity and Gauss--Bonnet term in
brane world scenarios}, Nucl. Phys. B {\bf 585}, 395 (2000).

\item A. Mazumdar and J. Wang, {\it A note on brane inflation}, Phys.
Lett. B {\bf 490}, 251 (2000).

\item M. Mintchev and L. Pilo, {\it Localization of quantum field on
branes}, Nucl. Phys. B {\bf 592}, 219 (2000).

\item W. M\" uck, K. S. Viswanathan and I. Volovich, {\it Geodesics and
Newton's law in brane backgrounds}, Phys. Rev. D {\bf 62}, 105019 (2000).

\item S. Mukohyama, {\it Brane world solutions, standard cosmology, and
dark radiation}, Phys. Lett. B {\bf 473}, 241 (2000).

\item S. Mukohyama, {\it Global structure of exact cosmological solutions
in the brane world}, Phys. Rev. D {\bf 62}, 024028 (2000); Erratum, ibid
{\bf 63}, 029901 (2001).

\item Y. S. Myung, G. Kang and H. W. Lee, {\it Randall--Sundrum choice in
the brane world}, Phys. Lett. B {\bf 478}, 294 (2000).

\item S. Nam, {\it Modeling a network of brane worlds}, JHEP 03, 005
(2000).

\item S. Nam, {\it Mass gap in Kaluza--Klein spectrum in a network of
brane worlds}, JHEP 04, 002 (2000).

\item I. P. Neupane, {\it Consistency of giher derivative gravity in the
brane background}, JHEP 09, 040 (2000).

\item S. Nojiri and S. D. Odintsov, {\it Brane world inflationinduced by
quantum effects}, Phys. Lett. B {\bf 484}, 119 (2000).

\item S. Nojiri and S. D. Odintsov, {\it Can we live on the brane in
Schwarzschild--anti de Sitter black hole?}, Phys. Lett. B {\bf 493}, 153
(2000).

\item T. Ozeki and N. Shimoyana, {\it Four--dimensional Planck scale is
not universal in fifth dimension in Randrall--Sundrum scenario}, Prog.
Theor. Phys. {\bf 103}, 1227 (2000).

\item S. K. Rama, {\it Brane world scenario with $m$--form field:
stabilisation of radion modulus and sel--tuning solutions}, Phys. Lett. B
{\bf 495}, 176 (2000).

\item S. Randjbar--Daemi and M. Shaposhnikov, {\it On some new warped
brane world solutions in higher dimensions}, Phys. Lett. B {\bf 491}, 329
(2000).

\item M. Sasaki, T. Shiromizu and K.--I. Maeda, {\it Gravity, stability
and energy conservation on the Randall--Sundrum brane world}, Phys. Rev.
D {\bf 62}, 024008 (2000).

\item H. Stoica, S. H. Henry Tye and I. Wasserman, {\it Cosmology in the
Randall--Sundrum brane world scenario}, Phys. Lett. B {\bf 482}, 205 (2000).

\item T. Tanaka and X. Montes, {\it Gravity in the brane world for
two--branes model with stabilized modulus}, Nucl. Phys. B {\bf 582}, 259
(2000).

\item T. Tanaka, {\it Asymptotic behavior of perturbations in
Randall--Sundrum brane world}, Prog. Theor. Phys. {\bf 104}, 545 (2000).

\item D. J. Toms, {\it Quantised bulk fields in the Randall--Sundrum
compactification model}, Phys. Lett. B {\bf 484}, 149 (2000).

\item D. Youm, {\it Solitons in brane worlds. I, II}, Nucl. Phys. B {\bf
576}, 106, 139 (2000).

\item D. Youm, {\it Probing solitons in brane worlds}, Nucl. Phys. B {\bf
576}, 123 (2000).

\item D. Youm, {\it Extra force in brane worlds}, Phys. Rev. D {\bf 62},
084002 (2000).

\item D. Youm, {\it Dilatonic $p$--branes and brane worlds}, Phys. Rev. D
{\bf 62}, 084007 (2000).


\item Z. Kakushadze, {\it Gravity in Randall--Sundrum brane world
revisited}, Phys. Lett. B {\bf 497}, 125 (2001).

\end{enumerate}

\end{document}